\documentclass[preprint,showpacs,printnumbers,amsmath,amssymb]{revtex4}
\usepackage{graphicx}% Include figure files
\usepackage{dcolumn}% Align table columns on decimal point
\usepackage{bm}% bold math
\usepackage{epsfig}
\begin{document}

\preprint{PREPRINT}

\title{Interplay between structure and density anomaly for an 
isotropic core-softened ramp-like potential}

\author{ Alan Barros de Oliveira}
\affiliation{ Universidade Federal do Rio Grande do Sul, Caixa
Postal 15051, 91501-970, Porto Alegre, RS, Brazil}
\email{oliveira@if.ufrgs.br}

\author{Paulo A. Netz}
\affiliation{ Instituto de Qu\'{\i}mica, Universidade Federal do Rio
Grande do Sul,
91501-970, Porto Alegre, RS, Brazil}

\author{ Marcia C. Barbosa}
\affiliation{ Universidade Federal do Rio Grande do Sul, Caixa
Postal 15051, 91501-970, Porto Alegre, RS, Brazil.}

\date{\today}

\begin{abstract}

Using molecular dynamics simulations and integral equations 
we investigate the structure, the thermodynamics and the dynamics of a
system of particles interacting  through a  continuous 
core-softened ramp-like interparticle potential. 
We found density, dynamic and structural anomalies similar to that 
found in water. Analysis of the radial distribution function for 
several temperatures at fixed densities show a pattern that 
may be related to the origin of density anomaly.
%We found that the density, at constant pressure, 
%has a maximum for a  certain temperature. The line 
%of temperatures of maximum density (TMD) was determined 
%in the pressure-temperature
%phase diagram. Similarly the diffusion coefficient at 
%a constant temperature, $D$, 
%has a maximum at a density $\rho_{max}$ and a minimum at a density 
%$\rho_{min}<\rho_{max}$. In the pressure-temperature phase-diagram
%the line of extrema in diffusivity is outside of TMD line. 
%The translational order parameter, $t$ and the 
%orientational order parameter, $Q_6$ also behave anomalously
%in a region of the phase-diagram which englobes both
%the dynamic and the thermodynamic anomalies regions,
%the same hierarchy of that one observed for the SPC/E water.

\end{abstract}

%\pacs{64.70.Pf, 82.70.Dd, 83.10.Rs, 61.20.Ja}
\maketitle

%%%%%%%%%%%%%%%%%%%%%%%%%%%%%%%%%%%%%%%%%%%%%%%%%%%%%%%%%%%
\section{Introduction}
%%%%%%%%%%%%%%%%%%%%%%%%%%%%%%%%%%%%%%%%%%%%%%%%%%%%%%%%%%%

Water is an anomalous substance in many respects.
Its specific volume at ambient pressure starts to increase 
when cooled below $T=4 ^oC$.  This density anomaly 
can be well explained by the tetrahedral structure of
water. Each molecule form hidrogen bonds with neighbors 
molecules by donating and receiving electrons from
the hidrogens. But this is 
not the only peculiarity of
water.  While for most
materials diffusivity decreases with increasing pressure, 
liquid water has an opposite behavior in a large region
of the pressure-temperature  phase diagram 
\cite{Er01,Ne01,Ne02a,Ne02b,Ne02}. %St99,Sc91
This diffusivity (or dynamic) 
anomaly is due to the following
mechanism: the increase in pressure disturbs the tetrahedral 
structure of water  by the  inclusion
of an  interstitial fifth molecule that shares an hydrogen bond with
another neighbor oxigen. As a result, the bond is weakened
and the molecule is free to move. The shared bond breaks
and the molecule, by means of a small rotation, connects to
another molecule enabling the translational diffusion  \cite{Ne02a}.
The structure and anomalies are therefore deeply related.

The quantification of structure  usually employs 
Errington and Debenedetti's translational order parameter \cite{Er01} $t,$ that
measures the tendency of pairs of molecules to adopt
preferential separations, and  Steinhardt's \cite{St83} orientational
order parameter $Q_6$. 
For normal liquids, $t$ and $Q_6$ increase upon compression, because
the system tends to be more structured. For 
systems with tetragonal symmetry \cite{Er01} the suitable
orientational order parameter is $q,$ that
quantifies the extend to which a molecule and its four
nearest neighbors assume a tetrahedral arrangement.
It was found that in SPC/E water
both $t$ and $q$ decrease upon compression in a certain region
of the pressure-temperature (P-T) phase diagram \cite{Er01}.
This region is referred as the region of structural anomalies.

Is the tetrahedral structure the only
one where anomalies would exist? The answer to 
this questions is no. Isotropic models 
are the simplest framework to understand the physics of
liquid state anomalies.  A number of such models, in which 
single component systems of particles interact via 
core-softened potentials \cite{pabloreview} have been proposed. 
They possess a 
repulsive core that exhibits a region of 
softening where the slope changes dramatically. This region can 
be a shoulder or a ramp. These isotropic models are design to
represent interacions in water and other materials in an effective way.

Recently, we investigated a system of particles interacting through a
core-softened, ramp-like potential\cite{Ol06a,Ol06b} that
has density, diffusion, and structural anomalies similar to
that found for SPC/E water. 
In this work, we use the same model to investigate the relation
between the local structure of the fluid, as measured by the
pair distibution function, and the density anomaly.

%%%%%%%%%%%%%%%%%%%%%%%%%%%%%%%%%%%%%%%%%%%%%%%%%%%%%%%%%%%
\section{The results}
%%%%%%%%%%%%%%%%%%%%%%%%%%%%%%%%%%%%%%%%%%%%%%%%%%%%%%%%%%%

\begin{figure}[ht]
\center
\includegraphics[clip=true,scale=0.6]{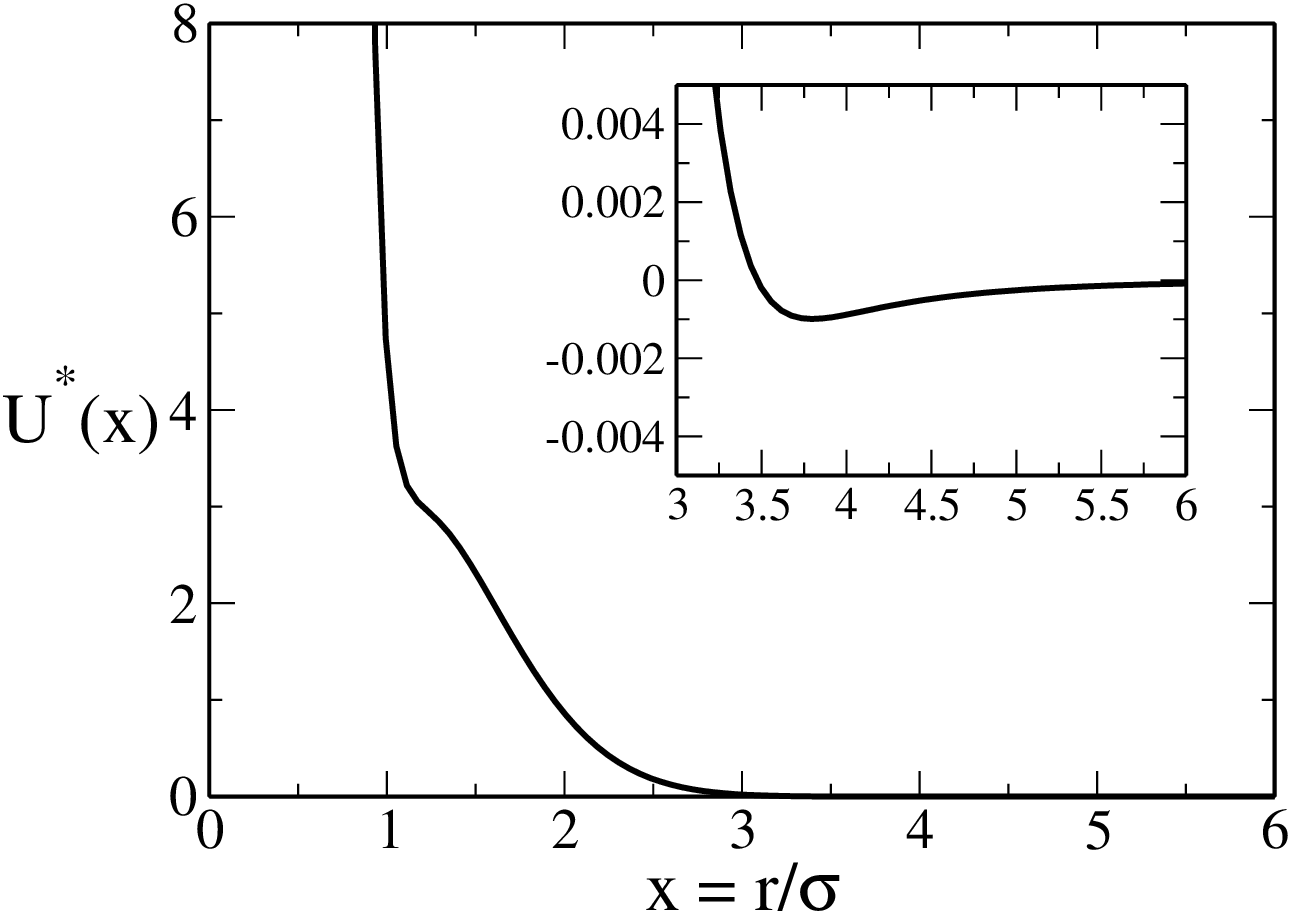}
\caption{Interaction potential from Eq. (\ref{eq:potential}) 
with parameters $a = 5,$ $r_{0}/\sigma = 0.7,$ and $c = 1,$ in
reduced units. The inset shows a zoom in the very small 
attractive part of the potential.
\label{cap:pot}}
\end{figure}

The model we study consists of a system of
$N$ particles  of diameter $\sigma$, inside
a cubic box whose volume
is $V$, resulting in a density number 
$\rho = N/V$. The  interacting 
effective potential between particles is given by

%%%%%%%%%%%%%%%%%%%%%%%%%%%%%%%%%%%%%%%%%%%%%%%%%%%%%%%%%%%
\begin{equation}
U^{*}(r)=4\left[\left(\frac{\sigma}{r}\right)^{12}-
\left(\frac{\sigma}{r}\right)^{6}\right]+
a\exp\left[-\frac{1}{c^{2}}\left(\frac{r-r_{0}}{\sigma}
\right)^{2}\right],
\label{eq:potential}
\end{equation}
%%%%%%%%%%%%%%%%%%%%%%%%%%%%%%%%%%%%%%%%%%%%%%%%%%%%%%%%%%

\noindent where $U^{*}(r)=U(r)/\epsilon.$
The first term of Eq. (\ref{eq:potential}) is a Lennard-Jones
potential of well depth $\epsilon$ and the second
term is a Gaussian centered on radius $r=r_{0}$ with height $a$
and width $c$.
With  $a=5,$ $r_{0}/\sigma=0.7$ and $c=1$, this 
potential  has  two length scales within 
a repulsive ramp followed by a very small attractive well,
as we can see in Figure \ref{cap:pot}.

%%%%%%%%%%%%%%%%%%%%%%%%%%%%%%%%%%
%%%%%%%%%%%%%%%%%%%%%%%%%%%%%%%%%%

The dimensionless pressure, $P^*,$ temperature, $T^*,$ and density, $\rho^*,$
are given in units of  $\sigma^3/\epsilon$, $k_{B} /\epsilon,$
and $\sigma^3,$ respectively. Here  $k_{B}$ stands for the 
Boltzmann constant.

Using integral equation with the Rogers-Young Closure \cite{RY84}
it is possible to analyse quickly the entire phase diagram.
For the model Eq. (\ref{eq:potential}) was found density anomaly \cite{Ol06a}.
In the neighborhood of the density
anomaly region, molecular dynamics simulations were
carried out, in order to investigate density,
diffusion, and structural anomalies.
Besides the confirmation of the presence of density anomaly
we demonstrated that
diffusion and structual anomalies are also
present in our model \cite{Ol06a,Ol06b}.

The relation between the several anomalies presented for the potential
Eq. (\ref{eq:potential})
is shown in Fig. 7 of Ref. \cite{Ol06b}. 
We found that 
the structural anomalous region have inside the
diffusion anomaly region which in its turns englobes the density anomaly
region. These hierarchy of anomalies 
is the same as the one found for the SPC/E water
(compare Fig. 7 of reference \cite{Ol06b} and
Fig. 4 of Ref. \cite{Er01}). 

%The TMD line indicates the region of thermodynamic anomaly region, inside
%which the density increases when the system is heated at constant pressure.
%The DE lines determinate the region of dynamic anomaly.
%Inside this region, diffusivity increases with increasing density.
%Between the curves of $Q_6$ maxima (B) and $t$ minima (A), 
%both order parameters decrease with increasing density (the 
%structural anomalous region).  These cascade
%of anomalies presents the same hierarchy as observed for the SPC/E
%water \cite{Er01,Ne01}, that is,
%density anomaly region inside the diffusion anomaly region,
%which is in its turn inside the structural anomaly region. 

%%%%%%%%%%%%%%%%%%%%%%%%%%%%%%%%%%
%\begin{figure}[ht]
%\center
%\includegraphics[clip=true,scale=0.6]{fig6.eps}
%\caption{Cascade of anomalies displayed by 
%our model.  See the text for more details.
%The region between curve B and curve A is the 
%structural anomalous region. 
%The diffusion extrema (DE) lines enclose the region
%inside which the diffusion decreases with density, and 
%the temperature of maximum density (TMD) line englobes the region that
%density anomaly appears.  
%\label{cap:cascade}}
%\end{figure}
%%%%%%%%%%%%%%%%%%%%%%%%%%%%%%%%%%

%\section{Density anomaly analysis}

\begin{figure}[ht]
\center
\includegraphics[clip=true,scale=0.6]{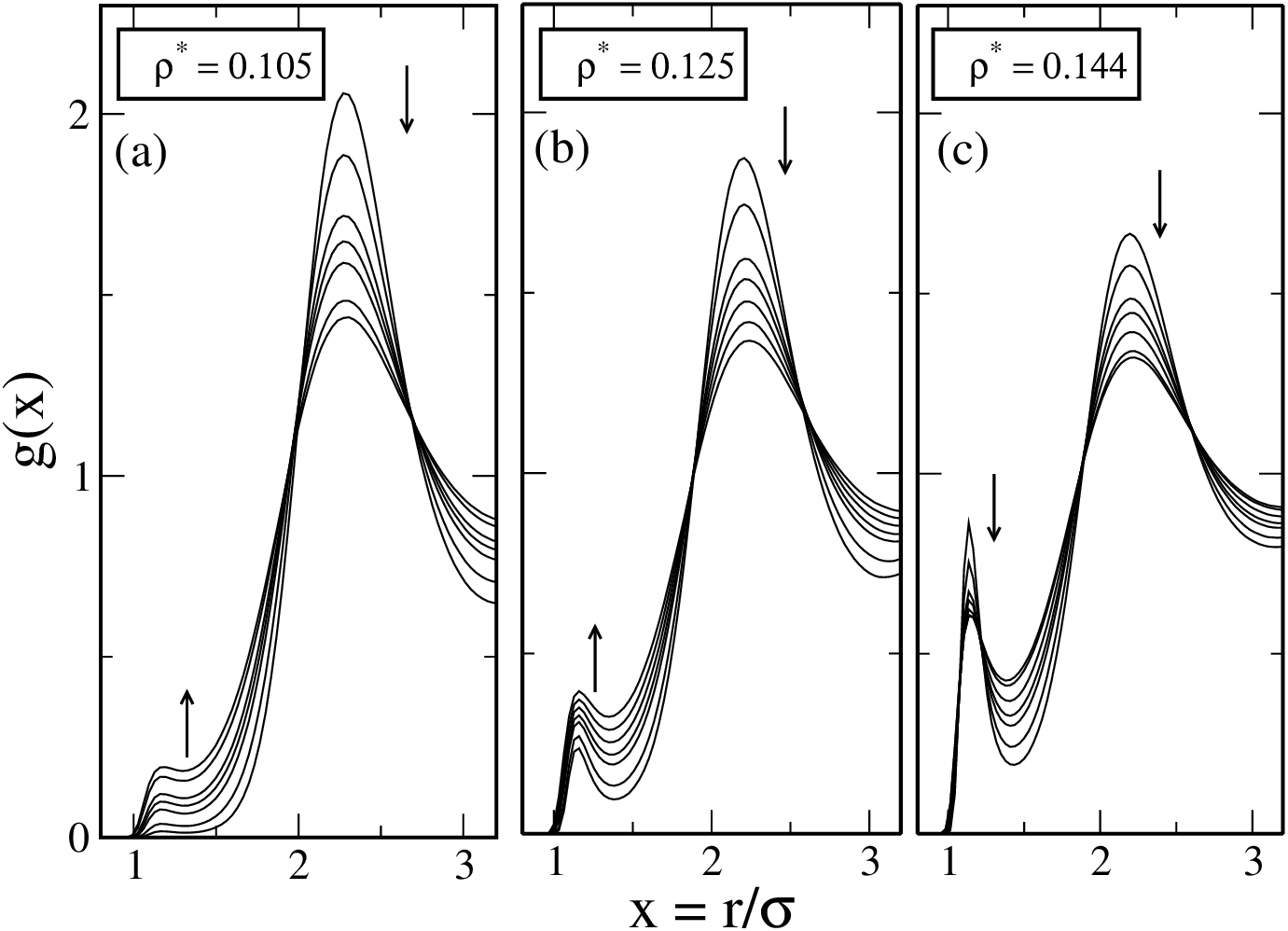}
\caption{Pair distribution functions for three
densities and seven temperatures of our model
Eq.(\ref{eq:potential}). The densities
are (a) $\rho^{*} = 0.105,$ (b) $\rho^{*}= 0.125,$ and
(c) $\rho^{*} = 0.144.$ These densities
are the same as those illustrated in Fig. \ref{cap:pt}.
The temperatures in
(a), (b), and (c) are $T^{*} = 0.15, 0.18,$
$0.23, 0.262,$ $0.3,$ $0.35,$ and $0.4$. The
arrows indicate the direction of temperature
growth, similar to the arrows in  Fig. \ref{cap:pt}.
\label{cap:allgrs}}
\end{figure}

\begin{figure}[ht]
\center
\includegraphics[clip=true,scale=0.6]{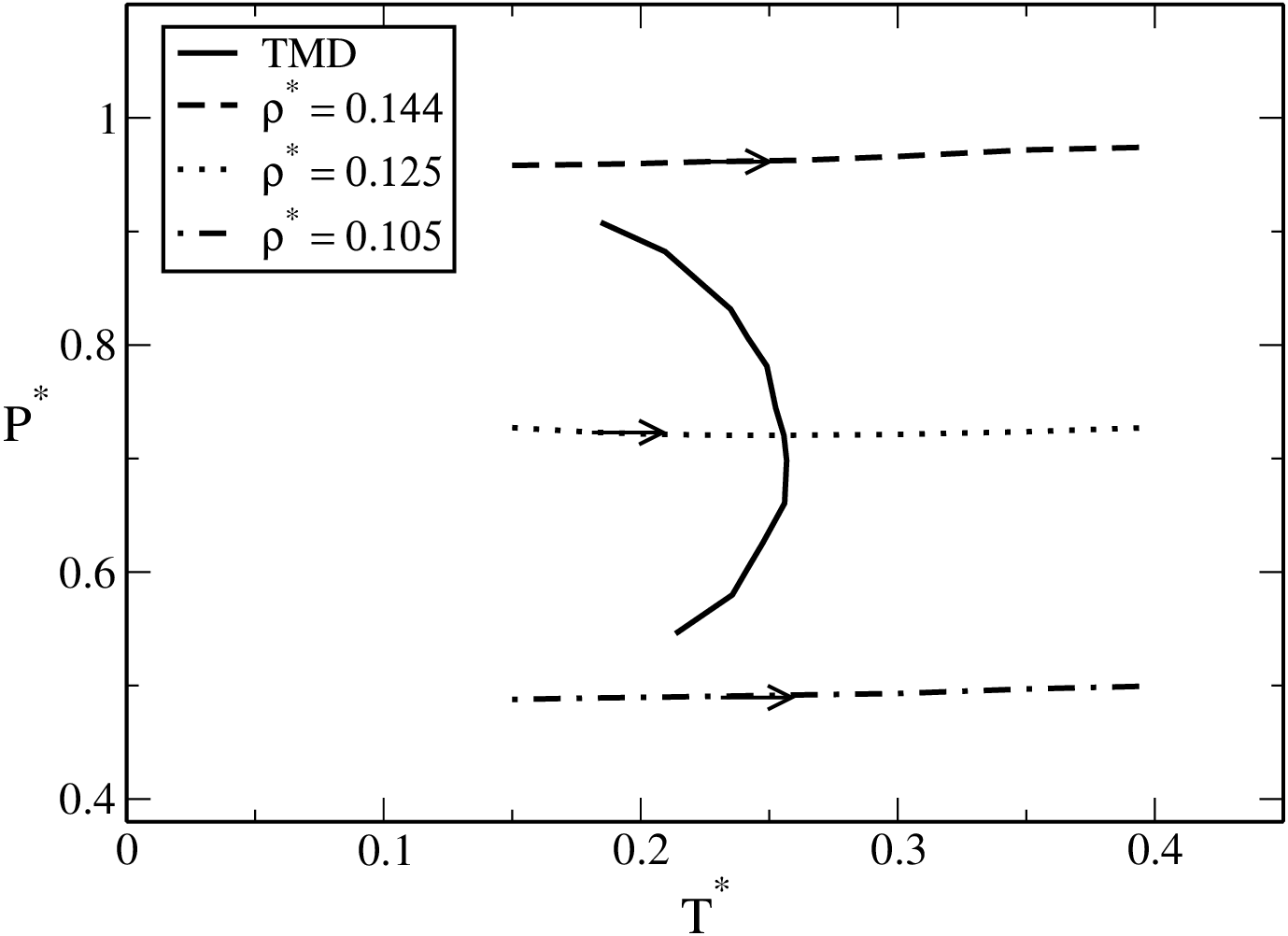}
\caption{Pressure-temperature phase diagram for the model
Eq. (\ref{eq:potential}). The three isochores shown in this figure
correspond to the regions below, inside, and above the region
where density anomaly occurs, i.e.
inside the TMD line (solid line).
The arrows indicate the temperature
growth and help the analysis in Fig. \ref{cap:allgrs}.
\label{cap:pt}}
\end{figure}

In order to understand the relation between structure and density anomaly, we analyse
the pair distribution functions of our model (Fig. \ref{cap:allgrs}). We see 
clearly three well defined regimes, namely:

(i) For a low density, below the region where density anomaly occurs
(see Fig. \ref{cap:pt}), the remarkable point is that 
the first peak (close to the core) of the $g(r)$ are too modest, and
the population of particles at this distance is negligible.
Note the arrows indicating the direction
of temperature increase. 

(ii) For intermediate density, inside the region of density anomaly,
we have a considerable increasing of the first peak of the $g(r)$ 
compared to the low density regime,
specially at low temperatures. Note that for temperatures
$T^{*} = 0.15, 0.18,$ and $0.23$,
just inside the density anomaly domain (see the TMD line in Fig. \ref{cap:pt})
the $g(r)$ had experience a special accented growth 
compared to the higher temperatures.
In the first peak the bottom $g(r)$ corresponds to
the lowest temperature and the top, to the highest temperature. The second peak
of the $g(r)$,
close to 2.5$\sigma$, has an opposite behaviour, and the arrow is downward.

(iii) For a high density, above the density anomaly region,
the interesting thing to note is the inversion of the arrow
in the first peak of $g(r)$. Close to the core, the  lowest temperature  has
the highest first peak. The trend in the second peak remains unchanged.

This analysis suggests that the behaviour of the 
$g(r)$ underlies the density anomaly effect in
a close way. The anomaly develops when the inner peak, close to core
distances, becomes increasingly important. 
We see that the  first
peak of the $g(r)$ tends to increase faster for low temperatures
upon compression than for the high temperatures. This suggests a
connection  between density anomaly and strucuture depending on
the derivative of the $g(r \approx \sigma)$  with respect to the temperature
at fixed density.
A similar study of $g(r)$ for several densities at constant temperature
was also applied to show its pattern related to the structural anomaly
region\cite{Ol06b}.

%%%%%%%%%%%%%%%%%%%%%%%%%%%%%%%%%

%%%%%%%%%%%%%%%%%%%%%%%%%%%%%%%%%

%%%%%%%%%%%%%%%%%%%%%%%%%%%%%%%%%%%%%%%%%%
\section{Conclusions}
%%%%%%%%%%%%%%%%%%%%%%%%%%%%%%%%%%%%%%%%%%

Using molecular dynamic simulations we have studied
the density behavior, the diffusivity, and the structure of fluids
interacting via a three-dimensional continuous core-softened
potential with a continuous force.
Our  model exhibits   a region of density anomaly, inside which
the density increases as the system is heated at constant pressure,
and a region of diffusion anomaly, where the diffusivity
increases with increasing  density \cite{Ol06a}. In the
pressure-temperature phase diagram, the density
anomaly region lies inside the diffusion anomaly one.
Complementary to the thermodynamic and dynamic
anomalies,  both $t$ and $Q_6$ behave anomalously
in a large region of the temperature$-$density plane.

This continuous core-softened pair potential, despite
not having long-ranged or directional interactions, exhibits
thermodynamic, dynamic \cite{Ol06a}, and structural anomalies\cite{Ol06b}
similar to the ones observed in SPC/E water \cite{Er01,Ne01}. Therefore, 
the presence of  anisotropy in
the interaction potential  is not a requirement for
the presence of thermodynamic, dynamic and structural
anomalies.

We also found  that the pattern of isochoric change
of the pair distribution fuction $g(r)$ with temperature
is closely related to the presence of density anomaly.
The inset of this anomaly may be related to the derivative
of population of 
molecules  with respect to the 
temperature in a distance corresponding to the core distance.

%%%%%%%%%%%%%%%%%%%%%%%%%%%%%%
\subsection*{Acknowledgments}
%%%%%%%%%%%%%%%%%%%%%%%%%%%%

We thank to 
the Brazilian science agencies CNPq, FINEP  and Fapergs 
for financial support.

\end{document}